\documentclass[rnote]{aa} 
\usepackage{graphicx}
\usepackage{txfonts}
\begin{document}
\newcommand{\mps}{\,m\,s$^{-1}$}
\newcommand{\kms}{\,km\,s$^{-1}$}
\newcommand{\msec}{\,m\,s$^{-1}$}

\title{Measuring differential rotation of the K-giant $\zeta$\,And
\thanks{Based on observations collected using the Bernard Lyot Telescope at Pic du Midi Observatory; the 1.93-m telescope at Haute-Provence Observatory, France; and the 8.2-m Kueyen telescope (VLT/UT2) of the European Southern Observatory, Chile (Prg. 081.D-0216(A)). }
 }


\author{Zs.~K\H{o}v\'ari\inst{1}
  \and H.~Korhonen\inst{2,3}
  \and L.~Kriskovics\inst{1}
  \and K.~Vida\inst{1}
  \and J.-F.~Donati\inst{4}
  \and H.~Le~Coroller\inst{5}
 \and J.~D.~Monnier\inst{6}
  \and E.~Pedretti\inst{7}
  \and P.~Petit\inst{4,8}
}

\offprints{Zs. K\H{o}v\'ari, \email{kovari@konkoly.hu}}

\institute{Konkoly Observatory of the Hungarian Academy of Sciences,
  H-1121, Konkoly Thege \'ut 15-17., Hungary\\
  \email{kovari@konkoly.hu}
  \and Niels Bohr Institute, University of Copenhagen, Juliane Maries Vej 30, DK-2100 K\o benhavn \O , Denmark
  \and Finnish Centre for Astronomy with ESO (FINCA), University of Turku, V{\"a}is{\"a}l{\"a}ntie 20, FI-21500 Piikki{\"o}, Finland
  \and IRAP-UMR 5277, CNRS \& Univ. de Toulouse, 14 Av. E. Belin, F-31400 Toulouse, France
  \and Observatoire de Haute-Provence, OHP/CNRS, F-04870 Saint-Michel l'Observatoire, France
  \and University of Michigan, 941 Dennison Building, 500 Church Street, Ann Arbor, MI 48109-1090.e, USA
  \and ESO, Karl-Schwarzschild Strasse 2, 85748 Garching bei M\"unchen, Germany
  \and LATT-UMR 5572, CNRS \& Univ. P. Sabatier, 14 Av. E. Belin, F-31400 Toulouse, France
}

\date{Received xxx xx, xxxx; accepted xxx xx, xxxx}

 
  \abstract
   {}
   {We investigate the temporal spot evolution of the K-giant component in the RS CVn-type binary system $\zeta$\,Andromedae to establish its surface differential rotation.} 
   {Doppler imaging is used to study three slightly overlapping spectroscopic datasets, obtained independently at three different observing sites. Each dataset covers one full stellar rotation with good phase coverage, and in total, results in a continuous coverage of almost three stellar rotations ($P_{\rm rot}=$17.8\,d). Therefore, these data are well suited for reconstructing surface temperature maps and studying temporal evolution in spot configurations. Surface differential rotation is measured by the means of cross-correlation of all the possible image pairs.}
   {The individual Doppler reconstructions well agree in the revealed spot pattern, recovering numerous low latitude spots with temperature contrasts of up to $\approx$1000\,K with respect to the unspotted photosphere,
   and also an asymmetric polar cap  which is diminishing with time. Our detailed cross-correlation study consistently indicate solar-type differential rotation with an average surface
shear $\alpha\approx0.055$, in agreement with former results. 
}
   {}



   \maketitle

%

\section{Introduction}

Stellar activity involves a range of magnetic activity phenomena, which are generated and modulated by the dynamo mechanism.
The differential rotation is of utmost importance in understanding the magnetic activity, since it is a key element of the stellar dynamo,
which has a controlling influence over the strength of magnetic fields generated, and thus that of the activity itself. The differential
rotation on stars with convective envelopes puts constraints on the large scale topology of the magnetic field, therefore giving important
information on the working of the dynamo beneath the surface. Nevertheless, on stars the detection of the surface differential rotation still
remains a challenging observational task, since stellar surfaces, except in rare cases (e.g., Gilliland~\& Dupree \cite{gil_dup},
Monnier et al. \cite{monnier07}), cannot yet be resolved directly. Thus, indirect surface reconstruction technique, like Doppler imaging
(see, e.g., Vogt et al. \cite{vogt87}, Piskunov et al. \cite{pisk90}), is needed.
Moreover, since starspots are commonly regarded as tracers of photospheric plasma motions, the pattern of surface differential rotation can be evolved
by investigating short-term spot redistribution through reconstructed time-series surface images
(e.g., Donati \& Collier Cameron \cite{doco97}, Weber \& Strassmeier \cite{west01}, K\H{o}v\'ari et al. \cite{kovetal04}, \cite{kovetal07a},
Barnes et al. \cite{baretal05}, Hussain et al. \cite{husetal06}, Weber \cite{web07}, etc.).

In this work the differential rotation of the K-giant component in the long-period  RS\,CVn binary $\zeta$\,Andromedae (hereafter $\zeta$\,And) is studied. $\zeta$\,And  (HD\,4502) is a single-lined
spectroscopic binary with a bright ($V=+4.1$ from SIMBAD) K-giant ellipsoidal variable component (Campbell et al. \cite{campetal1911}), and an unseen companion
of likely late-G or early-K dwarf (K\H{o}v\'ari et al. \cite{paper1}, hereafter Paper~I, see also the references therein). Although, its photometric variability is ruled
by the ellipticity effect, light curve modulation with the rotational cycle of 17.8 days is also present, suggesting spot activity (cf. Strassmeier et al. \cite{straetal89}).
This feature is in agreement with the strong Ca\,{\sc ii} $H$\&$K$ emission reported first by Joy \& Wilson (\cite{jowi49}). The star is known also
from its H$\alpha$, $UV$, and X-ray activity (cf. Eaton \cite{eato95}, Reimers \cite{reim80}, and Schrijver et al. \cite{schr84}, respectively).

The first extended study of the target using photometric modelling and Doppler imaging was presented in Paper~I. Doppler maps revealed spots
dominating at lower to medium latitudes with a typical temperature contrast of $\approx$750\,K,
as well as a weak and variable polar cap. Considering the distorted geometry, the fundamental system and stellar parameters were specified and an
investigation for differential rotation was presented. Recently, Korhonen et al. (\cite{paper2}, Paper~II from here on) measured the angular diameter of the star
with the Very Large Telescope Interferometer (VLTI/AMBER) and specified its absolute dimensions. A new Doppler image was also reconstructed
from the high quality spectra obtained simultaneously from the VLT/UVES instrument. The resulting image, similarly to the ones in Paper~I, recovered
mainly cool spots at lower latitudes, amongst with the most dominant features concentrating at quadrature positions, and also some polar features,
but with significantly weaker contrast.

A unique opportunity presented itself during August and September 2008, when the star was measured almost consecutively from
Pic du Midi Observatory (France), from Haute-Provence Observatory (France), and from the European Southern Observatory (Chile, see Paper~II),
providing three independent (slightly overlapping) spectroscopic datasets, each of them suitable to reconstruct one single Doppler image
using four favorable mapping lines. In Sect.~\ref{sect_obs} a brief review of the three datasets is given, then, in Sect.~\ref{sect_dop},
the results from Doppler imaging are presented. Afterwards, in Sect.~\ref{sect_ccf} the Doppler maps are compared with
each other by using our sophisticated cross-correlation technique to search for the surface differential rotation pattern. Finally,
the results are summarized and discussed in Sect.~\ref{sect_dis}.


\section{Spectroscopic observations}\label{sect_obs}

The spectroscopic data were collected at three different observing sites independently, between 13 August and 1 October, 2008. Each dataset
covers roughly one stellar rotation (16$-$18 days) and the consecutive sets overlap one another by 1$-$2 days. In all the three cases, the covered
wavelengths include the 6400\,\AA\ region commonly used for Doppler imaging. This contains, amongst others, the mapping
lines Fe\,{\sc i}-6411, Fe\,{\sc i}-6421, Fe\,{\sc i}-6430 and Ca\,{\sc i}-6439. The summary of the observations is given in Table~\ref{tab1}.

All the spectra are phased using the same ephemeris as in Paper~I, fitting the zero phase to the conjunction with the secondary in front:
\begin{equation}\label{eq1}
{\rm HJD} = 2\,449\,997.223\pm0.017 + 17.769426\pm0.000040\times E.
\end{equation}

\subsection{NARVAL spectra}

The first dataset in the time-series with altogether 48 spectra was obtained during 10 nights between 13--31 August at
Pic du Midi Observatory (France) using the 2-m Bernard Lyot Telescope, equipped with the NARVAL spectrograph (Auri\`ere \cite{narval}).
The spectrograph covers the wavelength range between 3700\,\AA\ and 10000\,\AA\  with a resolving power ($\lambda/\Delta\lambda$) of 60\,000.
Each observation consists of four 30\,s long exposures obtained immediately after each other. These exposures were combined
into one observation to increase the signal-to-noise ratio (S/N). This way 10 individual observations (one per night) from different rotational phases were obtained.
The information in Table~\ref{tab1} is for the combined observations. The reference HJD of the Doppler map calculated from the NARVAL spectra
in Sect.~\ref{sect_dop} (i.e., the mean HJD) is 2454702.088.

\subsection{SOPHIE spectra}

The second set consisting of 10 spectra were collected between 30 August and 15 September, 2008 with the SOPHIE \'echelle spectrograph (Perruchot et al. \cite{sophie})
at the 1.93-m telescope of Haute-Provence Observatory (France). The spectrograph has a wavelength range of 3872$-$6943\,\AA.
The observations were carried out in the high efficiency mode (HE) using 100\,$\mu$ fibre, which gives $\lambda/\Delta\lambda$ of 40\,000.
The exposure time was usually 120\,s, but in two cases longer times were used: on September 6 the exposure time was 950\,s
and on September 11 166\,s. The mean HJD of the SOPHIE observations is 2454715.177.

\subsection{UVES spectra}

The third set of observations were carried out at the European Southern Observatory (Chile) with UVES (UV-Visual Echelle Spectrograph,
Dekker et al. \cite{uves}) spectrograph mounted on 8.2-m Kueyen VLT unit telescope. The data were collected during ten nights between 13 September and 1 October, 2008.
In the observations the red arm standard wavelength setting of 600\,nm was used with image slicer \#3. This configuration gives
a resolving power ($\lambda/\Delta\lambda$) of 110\,000 and a wavelength coverage of 5000$-$7000\,\AA. For every observation
the three 8\,s exposures obtained immediately after each other were combined into a single, very high S/N spectrum. For further details see Paper~II.
Computing the average HJD we get 2454730.621.

\begin{table}[]
\caption[]{Observational records of our spectroscopic data.
}
\label{tab1}
\begin{flushleft}
 \begin{tabular}{ccccccc}
  \hline
  \noalign{\smallskip}
 HJD & phase\tablefootmark{a} & S/N  & date & instrument & $n$\tablefootmark{b} & $t_{\rm exp}$\tablefootmark{c} \\
2454000+   &  &   & (2008) &   &   & [s] \\
  \noalign{\smallskip}
  \hline
  \noalign{\smallskip}
691.5464 & 0.178 & 446 & 13 Aug & NARVAL & 4 & 30 \\
696.5723 & 0.461 & 258 & 18 Aug & NARVAL & 4 & 30 \\
699.6046 & 0.632 & 242 & 21 Aug & NARVAL & 4 & 30 \\
701.5989 & 0.744 & 679 & 23 Aug & NARVAL & 4 & 30 \\
702.5645 & 0.798 & 384 & 24 Aug & NARVAL & 4 & 30 \\
703.5135 & 0.852 & 270 & 25 Aug & NARVAL & 4 & 30 \\
704.5673 & 0.911 & 296 & 26 Aug & NARVAL & 4 & 30 \\
705.5435 & 0.966 & 329 & 27 Aug & NARVAL & 4 & 30 \\
708.6240 & 0.139 & 421 & 30 Aug & SOPHIE  & 1 &  120\\
708.6278 & 0.140 & 371 & 30 Aug & NARVAL & 4 & 30 \\
709.5420 & 0.191 & 445 & 31 Aug & SOPHIE  & 1 & 120 \\
709.6025 & 0.194 & 291 & 31 Aug & NARVAL & 4 & 30 \\
710.5015 & 0.245 & 335 & 01 Sep & SOPHIE  & 1 & 120\\
711.6389 & 0.309 & 441 & 02 Sep & SOPHIE  & 1 & 120\\
714.6815 & 0.480 & 261 & 05 Sep & SOPHIE  & 1 & 120\\
715.6242 & 0.533 & 198 & 06 Sep & SOPHIE  & 1 & 950\\
717.4957 & 0.639 & 269 & 08 Sep & SOPHIE  & 1 & 120\\
718.5432 & 0.698 & 441 & 09 Sep & SOPHIE  & 1 & 120\\
720.5707 & 0.812 & 237 & 11 Sep & SOPHIE  & 1 & 166\\
722.8167 & 0.940 & 866 & 13 Sep & UVES    & 3 & 8\\
724.5445 & 0.035 & 258 & 15 Sep & SOPHIE  & 1 & 120\\
724.7046 & 0.046 & 650 & 15 Sep & UVES    & 3 & 8\\
726.7101 & 0.159 & 586 & 17 Sep & UVES    & 3 & 8\\
727.6870 & 0.214 & 628 & 18 Sep & UVES    & 3 & 8\\
728.7387 & 0.273 & 762 & 19 Sep & UVES    & 3 & 8\\
730.7203 & 0.384 & 693 & 21 Sep & UVES    & 3 & 8\\
732.7287 & 0.497 & 914 & 23 Sep & UVES    & 3 & 8\\ 
734.6517 & 0.606 & 785 & 25 Sep & UVES    & 3 & 8\\ 
736.6936 & 0.721 & 612 & 27 Sep & UVES    & 3 & 8\\
740.7533 & 0.949 & 802 & 01 Oct & UVES    & 3 & 8\\
  \noalign{\smallskip}
 \hline
 \end{tabular}
\end{flushleft}
\tablefoot{
\tablefoottext{a} Phases computed using Eq.~\ref{eq1}.
\tablefoottext{b} Number of exposures combined into a final spectrum.
\tablefoottext{c} Lengths of the single exposures.}
\end{table}

\section{Doppler imaging with {\sc TempMap}$_\epsilon$}\label{sect_dop}

All of the three datasets cover roughly one stellar rotation, and thus allow to reconstruct one Doppler image for each set.
For the surface image reconstruction we started from the Doppler imaging code {\sc TempMap} (Rice et al. \cite{tempmap}) which,
afterwards, was extended by us ({\sc TempMap}$_\epsilon$, Paper~I) to take into account the non-spherical shape of stars
in close binary systems.
In our approximation, the real Roche-geometry is simplified to a rotation ellipsoid, that is elongated
towards the other component.
We note here that {\sc TempMap}$_\epsilon$ is the only Doppler imaging code available so far, that can use non-spherical surface geometry.
 We also note, that neglecting the ellipticity in the Doppler reconstruction would
yield a significant rise in the goodness-of-fit value of the line profile inversion (see Papers~I-II).
Eventually, the astrophysical input data of $\zeta$\,And, including the most feasible distortion parameter,
are adopted from Paper~I (Table~2 therein).

\begin{figure}
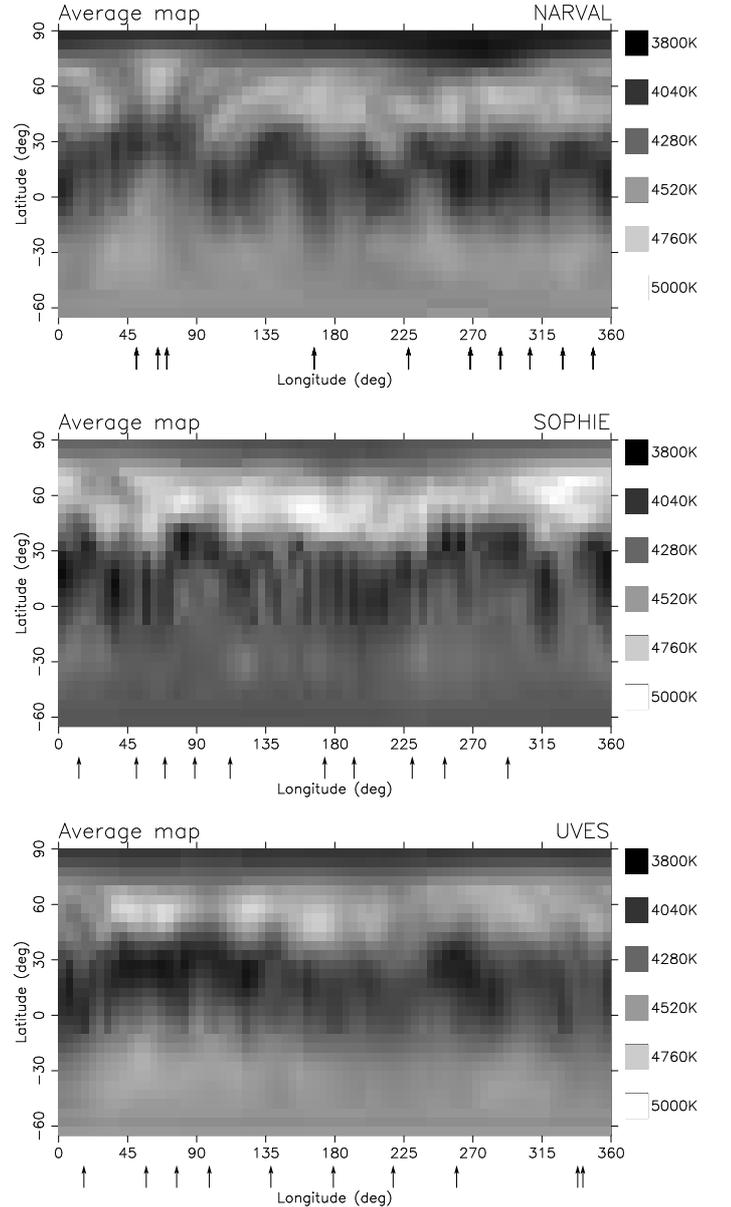

  \resizebox{\hsize}{!}{\includegraphics{average_narval.eps}}

  \vspace{0.3cm}

 \resizebox{\hsize}{!}{\includegraphics{average_sophie.eps}}

\vspace{0.3cm}
 
 \resizebox{\hsize}{!}{\includegraphics{average_uvnoalpha.eps}}

  \caption{From top to bottom: the combined (Fe\,{\sc i}-6411+Fe\,{\sc i}-6421+ Fe\,{\sc i}-6430+Ca\,{\sc i}-6439) Doppler maps in time series
  for the NARVAL, SOPHIE, and UVES data, respectively. Arrows underneath indicate the observational phases.}
  \label{fig_dop}
\end{figure}

Doppler imaging was performed for a sample of well known mapping lines, that usually produces the most reliable outcome.
In practice, we use only those lines that are jointly covered by all the three datasets: Fe\,{\sc i}-6411, Fe\,{\sc i}-6421, Fe\,{\sc i}-6430 and Ca\,{\sc i}-6439.
The resulting Doppler images are plotted in Fig.~\ref{fig_dop}.
Albeit a higher resolving power would allow for a larger amount of grid points to be used in the map, for the comparison reasons we apply the same grid size of $5\degr\times5\degr$ in all the maps.
The individual
reconstructions well agree in the revealed spot pattern, i.e., numerous low latitude spots are recovered with temperature contrasts of up to $\approx$1000\,K with
respect to the undisturbed photosphere of 4600\,K. In addition, an asymmetric cool polar feature is also present at around $\phi\approx0.75$, however,
switching from the NARVAL map to the SOPHIE and UVES maps, i.e., from August to September 2008 the contrast became weaker and the spot concentration
shifted backwards in phase. We refer also to Paper~II, wherein, using the UVES observations, a thorough Doppler imaging study was carried out.


\section{Cross-correlation study}\label{sect_ccf}

Surface differential rotation (DR) of a spotted star can be measured by cross-correlation of subsequent Doppler images (Donati \& Collier Cameron \cite{doco97}).
Nevertheless, even when having reliable Doppler images, short-term spot changes (e.g., vivid interaction between neighbouring spots, emerging a new
spot or disintegrating an old one, etc.) can easily mask the trace of the surface DR in the cross-correlation function (hereafter ccf) map. To emphasize the DR pattern,
and to diminish differences originating from individual spot motion and evolution, we use our newly developed cross-correlation technique
ACCORD (`Average Cross-CORrelation of time-series Doppler images', see e.g., in K\H{o}v\'ari et al. \cite{kovetal04}, \cite{kovetal07a}, Paper~I).
The basic idea of the method is, that averaging the more ccf maps, the better result can be achieved by boosting
such cross-correlation features that are jointly present, like the DR pattern itself.

\begin{figure}[t]
  \resizebox{0.9\hsize}{!}{\includegraphics{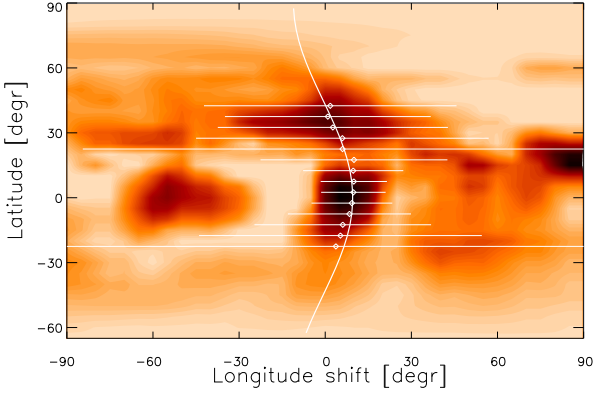}}
  \caption{Applying the ACCORD technique for the three available Doppler images plotted in Fig.~\ref{fig_dop}. Better correlation on the resulting ccf map is represented
  by darker shade. The average longitudinal cross-correlation functions
  in bins of 5\degr\ are fitted by Gaussian curves (Gaussian peaks are indicated by dots, the FWHMs by horizontal lines). The continuous line represents the
  best fit quadratic DR law.}
  \label{fig_ccf1}
\end{figure}

From the Doppler maps shown in Fig.~\ref{fig_dop} three cross-correlation pairs can be formed: NARVAL/SOPHIE (N/S), NARVAL/UVES (N/U), and SOPHIE/UVES (S/U).
Those ccf maps then can be averaged. However, a simple averaging will not definitely boost the DR pattern, since the shape of the correlation pattern depends on the
time gap between the paired Doppler images, i.e., the larger the time gap, the larger the elongation of the individual 1-D longitudinal cross-correlation functions
parallel to the equator (in practice, these are longitude strips in bins of 5\degr, see, e.g., K\H{o}v\'ari et al. \cite{kovetal04}). Thus, to avoid fuzziness,
a linear renormalization must be done before averaging. When the final average ccf map is assembled, the best-correlating pattern can be fitted by an
assumed form of the DR law. The result of applying ACCORD to the three Doppler images is plotted in Fig.~\ref{fig_ccf1}. For the latitude ($\beta$) dependent
rotation law we assume the usual quadratic form of
$\Omega(\beta)=\Omega_{\rm eq}\ (1-\alpha\sin^2\beta)$, where $\Omega_{\rm eq}$ is the equatorial angular velocity
and $\alpha=(\Omega_{\rm eq}-\Omega_{\rm pole})/ \Omega_{\rm eq}$ is the dimensionless surface shear parameter. 
The best fit yields solar-type equatorial acceleration:
\begin{equation}\label{eq2}
\Omega(\beta)=20.763\ (1-0.053 \sin^2\beta)  \, \, {\rm[\degr/d]}.
\end{equation}

\section{Discussions and conclusion}\label{sect_dis}

\begin{figure}[t]
  \resizebox{0.9\hsize}{!}{\includegraphics{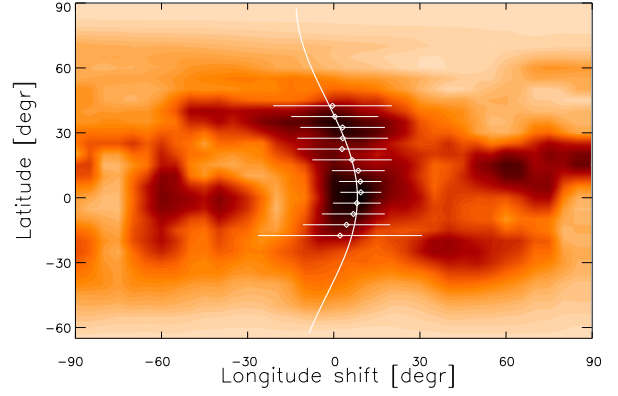}}
  \caption{Applying the ACCORD technique for the individual Doppler reconstructions using Fe\,{\sc i}-6411, Fe\,{\sc i}-6421, Fe\,{\sc i}-6430 and Ca\,{\sc i}-6439 mapping lines, separately.
  Otherwise as in Fig.~\ref{fig_ccf1}.}
  \label{fig_ccf2}
\end{figure}

The outcome of our new cross-correlation study seems to be fairly consistent with the result in Paper~I, where $\alpha$ of $0.049\pm0.003$ was derived.
To test the reliability of our method and result we repeat the correlation study, but instead of using the combined Doppler images in Fig.~\ref{fig_dop}, we start from
individual reconstructions, i.e., we use Doppler images that are recovered individually for a single mapping line. This way we have four
maps (Fe\,{\sc i}-6411, Fe\,{\sc i}-6421, Fe\,{\sc i}-6430 and Ca\,{\sc i}-6439) for each datasets (N, S, U), separately, thus 12 cross-correlations
can be achieved and finally averaged. We expect that this way the signal-to-noise of the DR pattern in the ultimate average ccf map will improve.
The result with the fitted DR pattern is shown in Fig.~\ref{fig_ccf2} yielding nearly the same DR law as of Eq.~\ref{eq2}:
\begin{equation}\label{eq3}
\Omega(\beta)=20.689\ (1-0.055 \sin^2\beta)  \, \, {\rm[\degr/d]},
\end{equation}
nevertheless, the relative height of the fitted `ridge' in the ccf landscape (i.e., the maxima of the gaussian peaks) has significantly ($\approx$40\% on average) risen,
compared to the one in Fig.~\ref{fig_ccf1}. We can also estimate the errors of the derived parameters given in Eq.~\ref{eq3} by carrying out the whole process,
but with using only three mapping lines from the available four, i.e., we perform four cross-correlation studies separately and estimate the error bars from the 1-$\sigma$ of the
obtained DR parameters, getting $\varepsilon(\Omega_{\rm eq})=\pm0.056$\,\degr/d and $\varepsilon(\alpha)=\pm0.0022$.
We consider this result as a proof of the reliability and robustness of the ACCORD technique. A comparative summary of the derived DR parameters is given in Table~\ref{tab_ccfs}.

\begin{table}[b]
\caption[]{Comparing the DR parameters
}
\label{tab_ccfs}
\begin{flushleft}
 \begin{tabular}{llll}
  \hline
  \noalign{\smallskip}
 $\Omega_{\rm eq}$ & $\alpha$ & LT\tablefootmark{a} & note \\
$\rm[\degr/d]$  &  &  [d]  &   \\
  \noalign{\smallskip}
  \hline
  \noalign{\smallskip}
 19.027 & 0.049 & 381 & from Paper~I \\
 20.763 & 0.053 & 332 & using combined images \\
 20.689 & 0.055 & 315 & using separate reconstructions \\
  \noalign{\smallskip}
 \hline
 \end{tabular}
\end{flushleft}
\tablefoot{
\tablefoottext{a} Time needed by the equator to lap the pole by one full period.
}
\end{table}

Theoretical calculations predict that across the H-R diagram DR becomes stronger with increasing rotation period and evolutionary status (Kitchatinov \& R{\"u}diger \cite{kitrued99}).
Still, $\zeta$\,And shows relatively weak surface differential rotation.
This could be explained by large starspots not following the stellar surface flows, as they are anchored
at different depth in the convection zone. Recent investigation by Korhonen \& Elstner (\cite{kor_els}) using dynamo models shows that
the much weaker DR than what is used in the model is recovered when having only the large-scale dynamo field. The real input DR law can only be recovered when the
model also includes small-scale fields. The question is whether or not large starspots can really be caused by small-scale magnetic fields.

According to Kitchatinov \& R\"udiger (\cite{kitrued04}) giant stars in close binary system, e.g., RS\,CVn-type binaries, can also show anti-solar differential rotation, where the equator
rotates slower than the poles. Such a DR law has been detected, for example, in $\sigma$\,Gem (K\H{o}v\'ari et al. \cite{kovetal07a}).
Kitchatinov \& R\"udiger (\cite{kitrued04}) attribute the anti-solar differential rotation to strong meridional flow, which could be caused by barocline driving due to large
thermal spots or by field forcing of a close companion. In accordance with this, a poleward meridional flow has been reported for $\sigma$\,Gem with an average
velocity of 350\kms\ (K\H{o}v\'ari et al. \cite{kovetal07a}). On the other hand, no sign of such a meridional flow is seen on $\zeta$\,And (Paper~I). This observed
difference between $\zeta$\,And and $\sigma$\,Gem can be rooted in the different system geometry and dynamics (K\H{o}v\'ari et al. \cite{iau282}). Distortion and tidal forces could account
for the converse DR in these two cool giants.

\begin{acknowledgements}
ZsK, LK and KV are grateful to the Hungarian Science
Research Program (OTKA) for support under the grant K-81421.
This work is supported by the ``Lend\"ulet" Young Researchers' Program of
the Hungarian Academy of Sciences.
\end{acknowledgements}

\end{document}